\documentclass[10pt,english]{article}

\usepackage{amsmath}
\usepackage{epsfig}
\usepackage{subfigure}

\usepackage{verbatim}
\usepackage{algorithmic}
\usepackage{array}
\usepackage[caption=false,font=footnotesize]{subfig}
\usepackage{fixltx2e}
\usepackage{dblfloatfix}
\usepackage{url}
\usepackage{qtree}
\newcommand{\codett}[1]{\text{\tt {#1}}}

\newcommand{\citeref}[1]{~\cite{#1}}

\newcommand{\eqnref}[1]{(\ref{#1})}
\newcommand{\figref}[1]{Figure~\ref{#1}}

\newcommand{\secref}[1]{Section~\ref{#1}}
\newcommand{\tableref}[1]{Table~\ref{#1}}

\newenvironment{texteq}{%
    \begin{equation}%
        \begin{minipage}{0.9\linewidth}%
}{%
        \end{minipage}%
    \end{equation}%
    \ignorespacesafterend%
}

\usepackage{varioref}
\usepackage[T1]{fontenc}
\usepackage{babel}

\newcommand{\longversion}[1]{#1}\newcommand{\shortversion}[1]{}

\usepackage{tikz}

\usepackage[absolute]{textpos}
\newcommand\copyrighttext{%
  \footnotesize 
 
  DISTRIBUTION A. Approved for public release: distribution
unlimited.

  \noindent
   \copyright Datanova Scientific 2016. 
}
\newcommand\copyrightnotice{%
\begin{textblock}{10}(3.5,14.5)
\copyrighttext

\end{textblock}
}

\begin{document}

\newcommand{\Author}[1]{#1}
\newcommand{\Address}[1]{#1}
\newcommand{\Email}[1]{#1}
\title{Implementing graph grammars for intelligence analysis in OCaml}
\author{\Author{Rod Moten}\\
         \Address{Datanova Scientfic, LLC}\\
         \Email{}
         \and
         \Author{Kemafor Anyanwu-Ogan}\\
         \Address{North Carolina State University}\\
         \Email{}
         \and
         \Author{Sahibi Miranshah}\\
         \Address{North Carolina State University}\\
         \Email{}
         \date{March 19, 2016}
       }
\maketitle

\begin{abstract}
We report on implementing graph grammars    for intelligence analysis in OCaml. Graph grammars are represented as elements of an algebraic data type in OCaml. In addition to algebraic data types, we use other concepts from functional programming languages to implement features of graph grammars. We use type checking to perform graph pattern matching. Graph transformations are defined as implicit coercions derived from structural subtyping proofs, subset types, lambda abstractions, and analytics. An analytic is a  general-purpose OCaml function   whose output is required to match a graph pattern described by an element of an algebraic data type. By using a strongly-typed language for representing graphs, we can  ensure graphs produced from a graph transformation will match a specific schema. This is a high priority requirement for intelligence analysis.     
\end{abstract}
\section{Introduction}

\copyrightnotice

Intelligence analysis is the process of analyzing known facts about entities, events, and their relationships in order to predict future events and  facts about entities.
There are many methods and practices for performing intelligence analysis.
In the past decade, \textit{link analysis}~\citeref{wolverton_law:_2003} has risen as a valuable method for performing intelligence analysis over large-scale heterogeneous data sets.
Link analysis is the process of identifying new relationships between entities and events based on existing facts about the entities and events.
In link analysis, the intelligence data is often represented as a graph.
Edges between nodes represent semantic relationships between nodes.
 Visually, an analysis performs link analysis by drawing a new edge between two nodes or  creating a new node by combining nodes. In other words, the analyst performs transforms on the existing graph to reflect the discovery of new information.  A program can use information about the topology or features of the graph to infer new information automatically. For example, an analytic could infer two person nodes are close associates if 85\% of their friends-of edges point to the same node. The newly discovered relationship could be represented as  a new close-associate-of edge between the two person nodes.

\textit{Object-based production} (OBP) is a methodology for treating intelligence data produced from multiple sources as objects in a common language \citeref{chandler_p._atwood_activity-based_2015}.
An object represents an entity or event. 
The objects could be extracted from data sources of various modalities, such as still imagery, motion imagery, audio, unstructured text, and structured data. 
Performing link analysis over objects usually requires representing the objects and the relationships between objects as a graph.

A graph created from objects represents a knowledge base of all known facts obtained from multiple sources of intelligence.
Usually an analyst only needs a portion of these facts when performing an investigation. 
We call the subgraph containing these facts the \textit{mission graph}. The mission graph may be a transformation of a subgraph of the main graph. The transformation often entails removing edges and attributes that aren't needed in the mission and condensing multiple relationships into a single relationship.

In this paper, we report on a graph database we created to support extraction of mission graphs using graph grammars. To represent  graph grammars, we created a strongly-typed language, called \textit{Flutes}, that we implemented in  OCaml. The terms in Flutes  represent graphs and the types represent specifications of  graphs with the same topology.  Graph grammar rules are defined  using subset types, lambda abstractions, and analytics. The body of a lambda abstraction is a Flutes term. An analytic is a lambda abstraction whose body can be any OCaml expression that evaluates to a Flutes term.

We demonstrate the value of using concepts from strongly-typed functional programming by showing the following. In \secref{flutes-section}, we show how we implemented a graph database  for OBP in OCaml.  In \secref{grammar-rules-in-flutes-section},  we show how constructs from Flutes are used to define graph grammar rules. In \secref{unification-in-flutes-section}, we show how  queries over large graphs can be performed efficiently by representing graphs  symbolically.

\section{Flutes DB}\label{flutes-section}
\begin{figure}
\centering

\subfigure[Relationships with cycles]{\includegraphics[height=0.5in,width=2in]{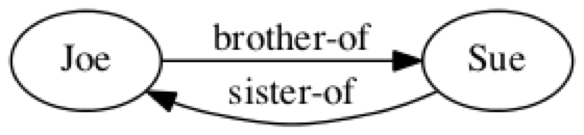}}\qquad
\subfigure[Relationships without cycles]{\includegraphics[height=1in,width=2in]{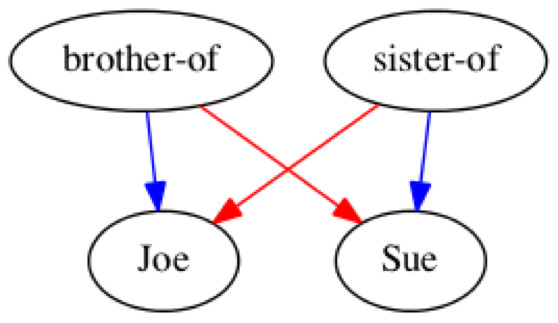}}
\caption{A cyclic relationship represented under the linked-object topology.}\label{siblings-graphs}
\end{figure}

Graphs used in OBP have a specific topology. We call the topology  the \textit{linked-object topology}. 
In the linked-object topology, objects are represented as DAGs.
 Relationships between objects are represented as DAGs where the root node is the name of the relation and the children are the related objects. 
A knowledge graph that has the linked-object topology does not have any cycles. Researchers who primarily work with RDF have trouble grasping this concept because cycles are common in RDF graphs. For example, consider the graph in \figref{siblings-graphs}a representing a man named Joe and a woman named Sue who are siblings. This is the common way to present relationships. The edge is used as the relation. As a result, inverse relations will form cycles.  Inverse relations can be represented without cycles using the linked-object topology as  depicted in \figref{siblings-graphs}b. In \figref{siblings-graphs}b, the blue edge and the red edge represent the first argument and the second argument of the relation, respectively.

Another feature of the linked-object topology not exploited by the RDF community is  classifying an edge as an \textit{essential property} or as an \textit{accidental property}\citeref{robertson_essential_2013}. An essential property represents a feature all objects of a specific type must have. However, the value of an essential property may be unknown. 
For example, birth date is an essential property of every person. 
 Accidental properties may exist for only some objects of a specific type. For example, having a brother is an accidental property of people.   

Since graphs with the linked-object topology do not have cycles, we can represent them as terms in a strongly-typed language. 
More specifically, an object  and its essential properties  can be represented as a record and accidental properties can be represented as $n$-ary predicates. We implemented such a language, which we call Flutes, in OCaml.   \figref{flutes-abstract-syntax} contains a portion of the definition of Flutes' abstract syntax as algebraic data type in OCaml.
We only show the definition of a fragment of the abstract syntax due to space limitations. \begin{figure}
\begin{verbatim}
type flutes_term =
  | Num of float | Str of string
  | Atom of Taxonomy.concept
  | Record of (Taxonomy.concept * flutes_term) list
  | List of flutes_term list
  | Bottom of Taxonomy.concept
  | Field_Selection of flutes_term * Taxonomy.concept
  | Var_Term of string
  | Term_Alias of string
  
and flutes_built_in_pred =
  | LessThan | LessThanEqualTo| GreaterThan
  | GreaterThanEqualTo | EqualTo
        
and flutes_prop_type =
  | Builtin_Pred of flutes_built_in_pred 
                   * flutes_term list
  | And of flutes_prop_type * flutes_prop_type
  | Or of flutes_prop_type * flutes_prop_type
  | Not of flutes_prop_type
  | Exists_Term of string * flutes_type 
                          * flutes_prop_type
  | TRUE | FALSE
  | InSequence of flutes_term* flutes_term list
        
and flutes_type=
  | NumTy | StrTy | ListTy of flutes_type
  | RecordTy of (Ontology.concept * flutes_type) list
  | EnumTy of Ontology.concept list
  | VoidTy
  | SubsetTy of flutes_term * flutes_type 
                            * flutes_prop_type
  | TyAlias of string
\end{verbatim}
\caption{Flutes terms and types defined in OCaml}\label{flutes-abstract-syntax}
\end{figure}

Objects are represented as records in Flutes, that is expressions created with the constructor \codett{Record}. Records are terms with a list of fields. The labels of the fields  
are concepts from a taxonomy. 
The module  \codett{Taxonomy}  encapsulates the taxonomy.
 The modules defines a method for creating a concepts from a string.  
\begin{verbatim}
             mk_concept : string -> concept
\end{verbatim}
\codett{Taxonomy} also maintains a total-ordering, an equivalence  relation and a join semi-lattice. The equivalence relation and the join semi-lattice can be used to define semantic relationships between concepts. For example, the equivalence relationship can be used to define synonyms of concepts  and the join semi-lattice can be used to define hypernyms and hyponyms of concepts. These relations are used in type subsumption, which we describe later in
the paper. \codett{Taxonomy} contains functions for adding pairs to the relations. We omit describing them in detail due to space limitation.

Predicates are implemented as records with special labels. These labels
indicate the argument position. For example, the logical expression $\mathrm{nickname}(\mathrm{``joe"},\mathrm{``joseph"})$ is defined as the record in \eqref{predicate-record-eg}.

\begin{texteq}\label{predicate-record-eg}
\begin{verbatim}
Record [("nickname", 
         [(Positional_Concept 0, Str "joe");
          (Positional_Concept 1, Str "joseph")]
        )]
\end{verbatim}
\end{texteq}

The elements constructed with \codett{Term\_Alias} are used to create references to terms.
In Flutes, terms can be assigned names.
These names are used at the arguments of \codett{Term\_Alias}. 

Assignment of names to terms is performed using a term declaration.
Term declarations are performed using the Flutes concrete syntax.
We created a parser for the  concrete syntax using menhir\longversion{\citeref{pottier_menhir_2007}}. 
Here's an example term definition.
\begin{verbatim}
   sue_grafton := {"name" : "Sue Grafton", 
                   "dob" : "1941-12-07",
                   "birth-place" = Kentucky};
\end{verbatim}

\noindent 
The expression on the left of \codett{:=} is the term name. 
  If a term definition is parsed successfully, it stores an S-expression of the term defined by the right of \codett{:=} in MongoDB \citeref{chodorow_mongodb:_2013}. The term name will be the key of the term in MongoDB. 

MongoDB is the persistent storage layer of Flutes DB.
Flutes DB is the combination of MongoDB, a library for the OCaml top-level and an OCaml standalone program for managing the Flutes terms in MongoDB. 
The library for the OCaml top-level environment contains a module named  \codett{Flutes\_cli}. This modules contains functions for defining Flutes terms and types and for retrieving terms and types from MongoDB. In addition, \codett{Flutes\_cli} contains functions for issuing commands to perform type checking, for applying lambda abstractions, and for running analytics. 

The \codett{Flutes\_cli} module hides the definition of the constructors of the algebraic data types in \figref{flutes-abstract-syntax}
to ensure Flutes terms and types are well--formed. \figref{flutes-cli} contains the
signatures of  some of these functions. For instance, the functions \codett{pred\_app}
and \codett{triple} construct records representing the application of a predicate to term. We can create the term in \eqnref{predicate-record-eg}  using \codett{triple} as follows.

\begin{verbatim}
      triple (str "joe") (str "joseph")
\end{verbatim}
The \codett{pred\_app} creates a predicate application term for predicates with various arity. We will use these functions when constructing Flutes terms and Flutes types in the rest of this paper instead of the constructors in \figref{flutes-abstract-syntax}. 

\begin{figure}
\begin{verbatim}
(******* FLUTES TERM FUNCTIONS **************)
type flutes_term
val str : string -> flutes_term
val num : int -> flutes_term           val num_f : float -> flutes_term
val list : flutes_term list -> flutes_term
val record : (string * flutes_term) list -> flutes_term
val atom : string -> flutes_term
val pred_app : string -> flutes_term list -> flutes_term
val triple : string -> flutes_term  -> flutes_term -> flutes_term
val record_select : flutes_term -> string -> flutes_term
val pred_arg_select : flutes_term -> int -> flutes_term
val var : string -> flutes_term
val term_name : string -> flutes_term
(******* FLUTES TYPE FUNCTIONS **************)
type flutes_type                       type flutes_prop_type
val str_ty : flutes_type        
val num_ty : flutes_type
val type_name : string -> flutes_type
val record_ty : (string  * flutes_type) list -> flutes_type
val pred_ty : string -> flutes_type list -> flutes_type
val triple_ty : string -> flutes_type * flutes_type  -> flutes_type
val subset_ty : flutes_term -> flutes_type -> flutes_prop_type
                                            -> flutes_type

(**INFIX OPERATORS FOR BUILT-IN PREDICATES *)
val ( === ) : flutes_term -> flutes_term -> flutes_prop_type                         
(***** INFIX OPERATORS FOR PROP TYPES ********)
(* and prop type *)
val ( ^^ ) : flutes_prop_type -> flutes_prop_type -> flutes_prop_type
(* or prop type *)
val ( ||| ) : flutes_prop_type -> flutes_prop_type -> flutes_prop_type
(* exists prop type *)
val ( ?? ) : string -> flutes_type -> flutes_prop_type -> flutes_prop_type
(******* FLUTES DB FUNCTIONS **************)
val abox_insert : string -> flutes_term -> unit
val find_members : unit -> unit
val mk_kb_class :string -> flutes_type -> unit
\end{verbatim}
\caption{Some of the signatures from \codett{Flutes\_cli}}\label{flutes-cli}
\end{figure}

Terms in MongoDB are partitioned into two collections, untyped terms and typed terms. Newly inserted terms are added to the untyped terms collection. Terms are inserted via the concrete syntax parser or using the function  \codett{abox\_insert} from the \codett{Flutes\_cli} module.  A term is moved to the typed terms collection if a Flutes type can be inferred from it. All well-formed Flutes terms have a Flutes type except those that reference terms that do not have a Flutes type.
 If a term alias refers to a term not in the typed terms collection, then the  term will not have a type.

In our current implementation of Flutes DB, type inference will only infer \textit{static types}. A static type is any  \texttt{flutes\_type} created with   any function in \figref{flutes-cli} other than \codett{type\_name} or \codett{subset\_ty}. Therefore, types created with the functions \codett{record\_ty} and \codett{pred\_ty} will be static types. Both of these functions use the  constructor \codett{RecordTy}. The main difference between the two is \codett{pred\_ty}\ creates a record type whose members are records representing predicate applications. The function \codett{triple\_ty} is a convenience version of  \codett{pred\_ty} . Types created with \codett{type\_name} represent references to types whose definitions are stored in MongoDB. Types whose definitions are stored in MongoDB are called \textit{classes}. Types created with \codett{subset\_ty} represent sets of terms that  satisfy a condition defined by a \codett{flutes\_prop\_ty}. We describe subset types in more detail in \secref{grammar-rules-in-flutes-section}.

 Determining membership of subset types and classes is performed by a separate task we call \textit{finding class members}.  Finding class members involves detecting which  typed terms belong to classes.  In Flutes DB, we create a collection in MongoDB for each class. Finding class members copies typed terms that belong to the class to the class' collection.    The class collections serve as indexes for graphs that have similar  topology and characteristics. Subset types represent sets of graphs whose  membership can change as new facts are added. For example, we use a subset type to represent people who have a financial transaction with the author Sue Grafton. 
As new financial transactions are added, new members may be added to this type. 

Finding class members is performed when a user evaluates the expression   \codett{find\_members()} in the top--level environment. Evaluating this expression causes the top-level environment to send  a message to running instances of a standalone OCaml program called \textit{Abox Manager}. The running instances of ABox Manager search the typed terms for members of classes.

We describe how \codett{find\_members} works via an example.  Suppose we evaluated the  code in \figref{person-transaction-code} in the top-level environment. The code in \figref{person-transaction-code} will result in four classes being defined: \codett{person}, \codett{trans}, \codett{orig\_of}, and \codett{recv\_of}. Running the parser on the concrete syntax in \figref{person-transaction-term-defs} will define a person named Sue and a person named Joe.  
\begin{figure}
\begin{verbatim}
joe := {"name"="Joe", "birth_date"="1984-06-27"};
sue := {"name"="Sue", "dob"="1941-12-07"};
t1 := {"amount" = 500.0, "type"=check()};
o1 := orig-of(joe, t1);
r1 := recv-of(sue, t2);
\end{verbatim}
\caption{Term definitions for two people involved in a financial transaction.}\label{person-transaction-term-defs}
\end{figure}
\noindent
In addition, it  defines a \$500 financial transaction involving Sue and Joe. All of these terms have Flutes static types, so ABox Manager  will  add all of these terms to the typed terms collection in MongoDB. Next, if the user evaluates \codett{find\_members\ ()} in the OCaml top-level, then ABox Manager instances will determine the classes each of the members belong to and add each member to the collection corresponding to the class in MongoDB. Both \codett{joe} and \codett{sue} would be added to   \codett{person}'s collection. Here is the  type inferred from \codett{joe}. 
\begin{texteq}\label{joes-type-eq}\begin{verbatim}
      record_ty [("name", str_ty); 
                 ("birth_date", str_ty)]
\end{verbatim}
\end{texteq}
This type is  a subtype of \codett{person} because we added  
\codett{dob} and
\codett{birth\_date} to the equivalence relation on the last line of the
code in \figref{person-transaction-code}. In Flutes, structural subtyping of record types use the equivalence relation and the lattice defined in the \codett{Taxonomy} module to pair fields. Normally, strict identity is used to pair fields.
\longversion{Under our definition of subtyping, it is possible for there
to be multiple proofs of the same subtyping judgment. We made our prover deterministic by requiring the fields of a record type to be ordered under the total ordering of labels defined in the module \codett{Taxonomy}. 
The prover finds pairs by looking to match the first field in the super type, with a field in the subtype. The prover scans the fields of the subtype in order. 
}
 
The term in the collection for \codett{person} bound to \codett{joe} has its \codett{birth\_date} renamed to \codett{dob}. 
In other words, \codett{joe} is the following term in \codett{person}.
\begin{verbatim}
  record [("name", "Joe"); ("dob", "1984-06-27")]
\end{verbatim} 
The result of proving a type is subtype of another type produces a coercion. When ABox Manager detects a term's inferred type is a subtype of a class, ABox Manager uses the coercion to convert the term to a term that belongs to the class. 
This is how type subsumption is implemented to support semantics relationship of attribute names.
\begin{figure}
\begin{verbatim}
let _ = mk_kb_class "person" 
        (record_ty [("name", str_ty); 
                    ("dob", str_ty)])
let _ = mk_kb_class "trans" 
        (record_ty 
             [("amount", str_ty); 
              ("type", enum_ty ["check", "cc"])])
                    
let _ = mk_kb_class "orig_of" 
        (triple_ty "orig-of" (type_name "person") 
                   (type_name "person")

let _ = mk_kb_class "recv_of 
        (triple_ty "recv-of" (type_name "person") 
                   (type_name "person")
                   
let _ = same_as "dob" "birth_date" /* last line */
\end{verbatim}
\caption{Code defining types for persons and financial transactions.}\label{person-transaction-code}
\end{figure}

\section{Grammar Rules in FlutesDB}\label{grammar-rules-in-flutes-section}
 Informally, a graph grammar production rule could be expressed as the logic formula $\forall g.P(g)\implies \exists g'.Q(g')$. 
In the formula,  $\forall g.P(g)$ is the left-hand side of the production rule and $\exists g'.Q(g')$ is the right-hand side of the production rule. The predicate $P$ defines the conditions a matching graph must have in order to apply the rule. 
The predicate $Q$ defines the conditions the output graph $g'$ will have after the rule has been applied to the matching graph.
The graph $g'$ is produced by performing some transformations to the graph containing $g$. 

\begin{figure}
\centering

\includegraphics[height=1.5in, width=5in]{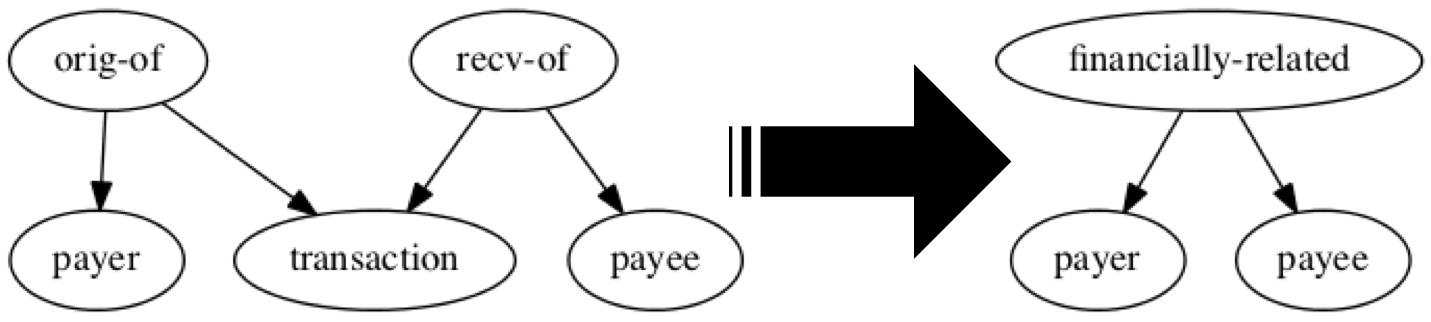}
\caption{Transformation of a financial transaction graph to a   financially related graph.}\label{financial-subgraphs}
\end{figure}

Flutes contains three constructs for representing grammar rules, subset types, lambda abstractions and analytics. Subset types are created with the function \codett{subset\_ty}. This function takes three arguments,  the binding term, the binding type, and the proposition type.
The first two arguments, the binding term and the binding type, determine the  the structure of the terms that belong to the subset type.
The third argument, the proposition type, indicates the properties each member of the subset type have.
The code in \eqref{mission-set-type-eq} contains an example of using a  subset type as a grammar rule. 
It defines a rule to transform a five-node four-edge subgraph into a three-node two-edge subgraph.  \figref{financial-subgraphs} contains a drawing of the transformation.
The code in \eqref{mission-set-type-eq} can be read logically as  make person $p$  financially
related to person $q$ if there exists a transaction $t$ where $p$ is the originator
of $t$ and $q$ is the receiver of $t$. Notice the binding term contains two variables, \codett{p} and \codett{q}, that occur
free in the proposition type.
\begin{texteq}\label{mission-set-type-eq}
\begin{verbatim}
let [p;q;r;s;t] = map var ["p";"q";"r";"s";"t"]
in
subset_ty 
  (triple "fi-related" p q) 
        ((triple_ty "fi-related" 
                (type_name "person",                                    
                 type_name "person"))) 
        ((??) "t" (type_name "trans")
           ((??) "s" (type_name "orig_of") 
             ((??) "r" (type_name "recv_of")
                (((triple "orig-of" p t
                       === s) 
                   ^^ 
                ((triple "recv-of" q t) 
                       === r)))))

\end{verbatim}
\end{texteq}
 
Finding members of the subset type involves finding terms that produce a
substitution that meets the following criteria. The application of the substitution to the binding term 
produce a term that belongs to the binding type and the application of the substitution to the proposition type produces a provable proposition.

Lambda abstractions and analytics are used to define graph grammars rules whose right-hand side use a functional graph transformation. These rules can be expressed informally as the logic formula  \(\forall g.S(g,I)\implies \exists g'.\kappa(g)=g' \wedge(S(f(g'),O)\). In this formula, $S$ is the predicate subsumed by. In other words, $S(x,C)$ means
$x$ is subsumed by type $C$ or class $C$ .  $f$ is the  graph transformation. $g'$ is the coerced term produced from the coercion  $\kappa$ derived from the proof of  the subtyping judgment of $g$'s inferred type and $I$.   In the case of lambda abstraction, $I$ is the class of the binding variable and $O$ is the type of the body, and $f(g')$ is the application of the lambda abstraction to $g'$. In the case of an analytic, f is an OCaml function of type \codett{flutes\_term} \codett{->} \codett{flutes\_term}. In this case, it is possible for $f(g')$ to not evaluate to a term subsumed by $O$. If this occurs, the grammar rule fails. 

The module \codett{Flutes\_cli} contains functions for defining grammar rules using lambda abstractions and analytics. Due to space limitation we omit the signatures and provide a short example of their use later in the paper.

\section{Large graphs in FlutesDB\\ }\label{unification-in-flutes-section}

We conducted experiments  to demonstrate FlutesDB can be used as a 5GL for mission extraction. In the experiments, we used subset types to extract a small low-dimensional mission set from a large high-dimensional data set. For the large high-dimensional data set, we used DBpedia \citeref{auer_dbpedia:_2007} and randomly generated financial data for people in DBpedia. The mission set are people who are financially related. 

We represent financially related people as terms of the form 
\begin{verbatim}
triple "fi-related" (term_name "x", term_name "y").
\end{verbatim}
These terms have to be constructed from collections of terms representing a financial transaction. The concrete syntax in \eqref{person-transaction-code} contains the definition of a financial transaction. The randomly generated terms for financial transactions may be incomplete. In other words, the transaction may  not have an originator or a receiver.
We did this to make the data set more realistic because intelligence data is often incomplete.

In addition to demonstrating mission set extraction, we show the ability to extract a mission target. 
A mission target represents  the objects an analyst is trying to discover using existing and newly collected intelligence data.
In our experiment, the newly collected intelligence data is represented as data added to FlutesDB after the mission set has been created.

To demonstrate our approach based on  FlutesDB is feasible  for mission extraction, we have to show it meets the following criteria.

\begin{enumerate}
\item\label{safety-criteria} An analyst can create a mission set and a mission target without having to write any special programming logic to ensure the transformation is safe.

\item\label{query-criteria} Queries and updates to the DB\ are performed without the analyst's knowledge. 

\item\label{performance-criteria} The amount of time to extract the mission set should be reasonable with respect to the size of the data set. 
\end{enumerate}

To meet Criteria \ref{safety-criteria} a solution  needs to prevent the execution of a transformation that will corrupt the integrity of data. 
Criteria \ref{query-criteria} means an analyst only specifies what they want in the mission set and the mission target opposed to specifying how to get the data using queries and API calls to retrieve it from the DB.

In Criteria~\ref{performance-criteria}, reasonable depends on the frequency of extracting mission sets. If an analyst
only needs to extract a mission set once a week, then two or three hours is
a reasonable amount of time to extract a mission set. 
If an analyst needs to extract several mission sets in a single eight-hour
work day, then a few minutes is a reasonable amount of time.

Detecting members of a mission target should always be fast.
Detecting whether a new term is a member of mission target should be on
the order of one millisecond for high-dimensional terms and ten microseconds
for low-dimensional terms. 
High-dimensional terms are objects such as people that may have 100s of attributes.
Low-dimensional terms are objects such as net-flow records \citeref{estan_building_2004} or tracks of a
moving vehicle \citeref{kreitmair_experimentation_2005} with fewer than 20 attributes.
This means if one million people are added, it should take less than 20 minutes 
to determine if those terms belong to a mission target.
This is suitable for investigations tracking the behaviors of people or other
entities with many features.  
If the new terms represent one million net-flow records or one million new
air tracks, it should take about 10 seconds to determine if they belong to a
mission target.
A typical PC  generates less than 1000 net-flow records per second
and GMTI\citeref{pannetier_gmti_2009} sensors produce less than 1000 tracks per second.
Therefore, 10 seconds is a reasonable amount of time for these cases.

For the experiment, we used  238,287,247 Flutes terms. These terms were created by converting   378,547,908 DBpedia triples into terms  in Flutes' concrete syntax and from  approximately 17 million  financial transaction terms for people in DBpedia. The graphs containing the transactions and the attributes of the people graphs have high dimensionality. \longversion{\figref{people-transaction-graph-full} illustrates the dimensionality pictorially.} 

 Conversion of  DBpedia was performed by adapting an algorithm for automatically detecting objects in an RDF\ graph. The objects detected are trees of depth one. We treated the edges as essential properties. Therefore each of these trees were converted into a record whose fields corresponded to the edges of the tree. Any property in the RDF graph not detected as an essential property we considered to be an accidental property. We represented these edges as predicate applications, that is terms created using the function \codett{pred\_app}.

\longversion{\begin{figure}
\centering
\includegraphics[height=4in,width=5in]{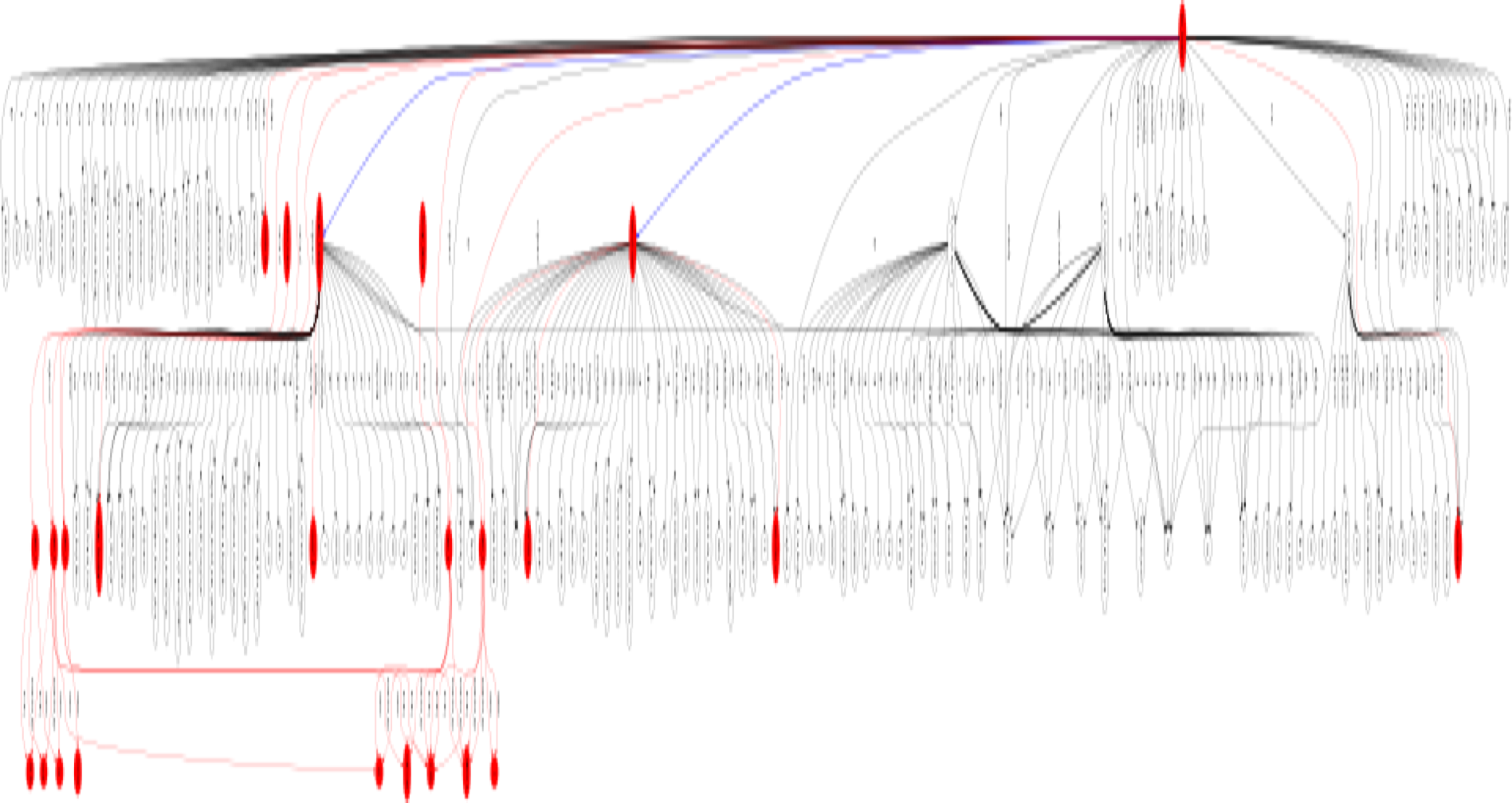}
\caption{Sue Grafton in financial transactions with two other people.}\label{people-transaction-graph-full}
\end{figure}}

For the experiment, we  made  the mission set   all financially related people. This type is defined in  \eqnref{mission-set-type-eq}. This mission set only needs to contain links between people who have at least one financial transaction in common. We don't need to include the financial transaction in the mission set. In addition, we don't need to include most of the attributes of a person. We only need the person's name and date of birth. \longversion{
\figref{sue-grafton-financially-related} graphically depicts a portion of the mission extracted from the graph in \figref{people-transaction-graph-full}.

\begin{figure}
\centering
\includegraphics[height=3in, width=5in]{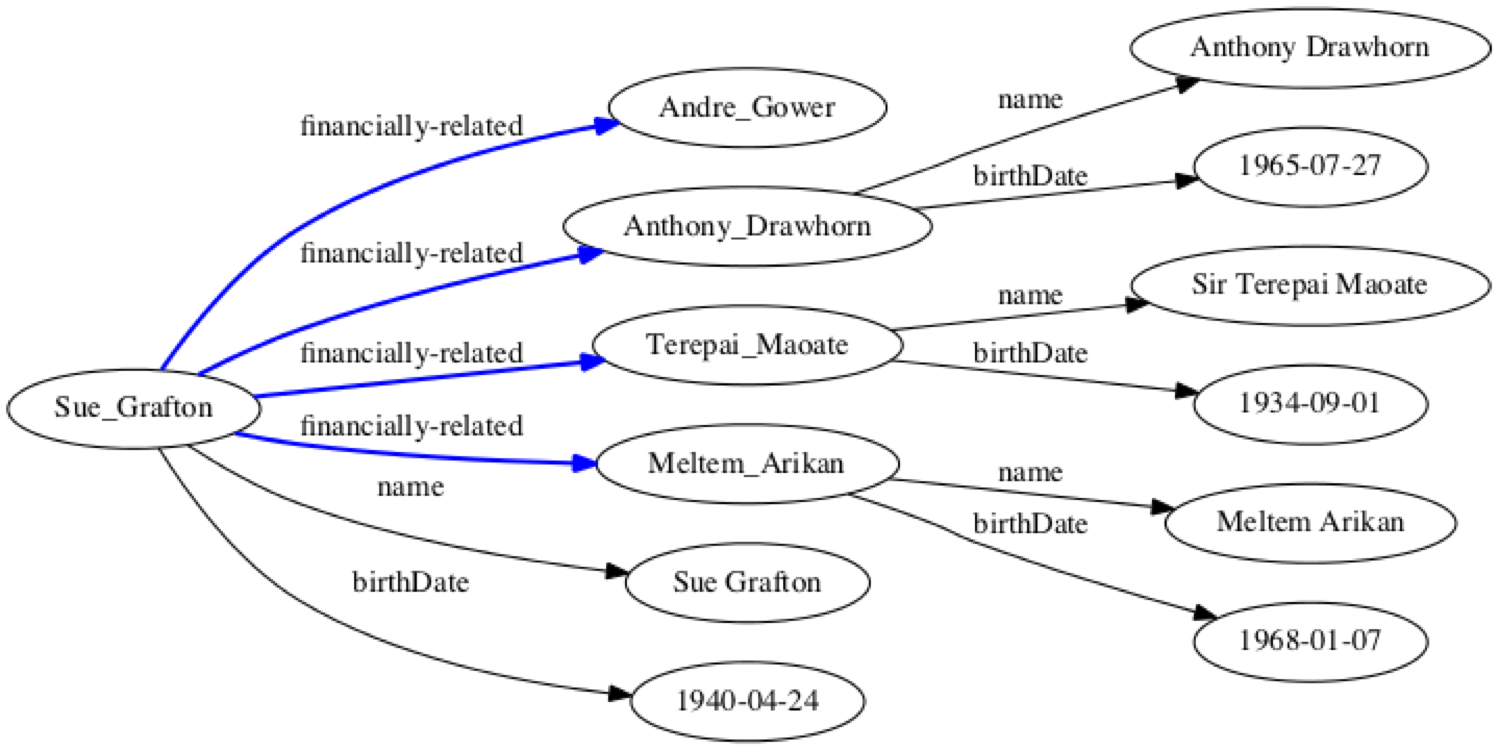}
\caption{People financially related to Sue Grafton from \figref{people-transaction-graph-full}.}\label{sue-grafton-financially-related}
\end{figure}
}

FlutesDB uses unification to find members of subset types.
FlutesDB creates a collection of disjunctive skolemized clauses from the subset type.
FlutesDB determines the skolem variables based on variables bound by the existential quantifier \texttt{??}. 
In  \eqref{mission-set-type-eq}, FlutesDB creates two skolem variables for \texttt{s} and \texttt{r}. 
FlutesDB scans the typed terms for instances to replace the skolem variables in the clauses.
Then it performs unification to determine bindings for the free variables in the binding term of the subset type.
In \eqref{mission-target-eq} the variables are \codett{p} and \codett{q}. If the substitution of the free variables produces a term that belongs to or is subsumed by the binding type, then the term or its coercion is made a member of the class represented by the subset type.

\begin{table*}
\centering
 \begin{tabular}{|l|c|}\hline
 \codett{find\_members} & elapsed time \\ \hline
orig\_of   &    24.53m \\
recv\_of    &    32.26m\\
fi\_related    &    45.23m\\
mission set (total)$^\mathrm{a}$ & 102.02m\\
m\_target (initial)$^\mathrm{b}$ &  6.23239m\\
m\_target (new. per term)$^\mathrm{c}$ & 1.17703s\\
\hline
\multicolumn{2}{l}{\footnotesize{$^\mathrm{a}$ Sum of finding members of
\codett{orig\_of}, \codett{recv\_of}, and \codett{fi\_related}.}}
\\
\multicolumn{2}{l}{\footnotesize{$^\mathrm{b}$ Performed over the existing
set of terms.}}
\\
\multicolumn{2}{l}{\footnotesize{$^\mathrm{c}$ Average for 5 new financial
transactions (20 terms). The elapsed}}
\\
\multicolumn{2}{l}{\footnotesize{\,\,\,\, time includes inferring types,
adding to the mission set, and adding}}
\\
\multicolumn{2}{l}{\footnotesize{\,\,\,\, to the mission target.}}
\end{tabular}
\caption{Average elapsed times on DBpedia with financial transactions}\label{elapsed-times-table}
\end{table*}

To make FlutesDB efficient, all the typed terms aren't scanned. 
Only a subset of the terms belonging to the type of the skolem variables are scanned. 
In the case of the mission set in \eqref{mission-set-type-eq}, we only scan terms from \codett{orig\_of} and \codett{recv\_of}.
However, we do scan  $n^m$ terms where $n$ is the number of terms in \codett{orig\_of} and $m$ is the number of terms in  \codett{recv\_of}.
When using a member of \codett{orig\_of} as a skolem variable, we only need to scan the set of members of \codett{recv\_of} that contains the same term of type \codett{trans} as the member of \codett{orig\_of}. 
FlutesDB creates an   adjacency list that maps each term to all the terms it contains. FlutesDB uses this adjacency list to create the set of terms to scan for members of recv\_of. This set only contains two terms for our data set.  Therefore, we only scan \(n^{2}\) terms. Once we find this transaction, unification will produce a substitution for the two arguments of \codett{fi-related}.

Before performing this process, FlutesDB has to find the members of \codett{orig\_of}
and \codett{recv\_of}.
FlutesDB detects the dependency of the mission set has on \codett{orig\_of} and \codett{recv\_of} when it creates the clauses from \eqnref{mission-set-type-eq}.
FlutesDB will only use newly added typed terms as candidates of \codett{orig\_of}
and \codett{recv\_of}.
\tableref{elapsed-times-table} shows the running times to find members of \codett{orig\_of}, \codett{recv\_of} and \codett{fi\_related}. 

In the experiment, we made finding people related to Sue Grafton the mission target. We define the mission target as the class \codett{m\_target} using the type in \eqnref{mission-target-eq}.

\begin{texteq}\label{mission-target-eq}
\begin{verbatim}
let target = term_name 
        "http://dbpedia.org/resource/sue_grafton"
in let p;f = var "p"; var"f"
in
subset_ty "p" (type_name "person") 
  ((??) "f" (type_name "fi_related")
    ((triple "fi-related" p target === f) 
    ||| 
    (triple "fi-related" target p === f)))
\end{verbatim}
\end{texteq}

\tableref{elapsed-times-table} also contains the elapsed
 time to find members of \codett{m\_target}. The elapsed time show that  in a  little more than one second, we can type five new financial transaction (20 terms), add new terms to the mission set, and add new terms to the target set. This is within the range specified by Criteria \ref{performance-criteria}. 

We can use an analytic to do something more dynamic than the subset type, such as determining membership of a class based on whether a term is near a member of a class of money laundering suspects. Finding the nearest neighbor needs the ability to traverse the adjacency list.
We have an OCaml module, called \codett{Graph\_api} for traversing the accidental properties of the graph.
 Due to space limitation, we don't provide it here.
 This module allows programmers to traverse the entire graph, so it supports the ability to write nearest neighbor and other graph algorithms.
 This module will also have a function to return the names of the classes a term belongs to.

Assume a programmer created a nearest neighbor function using the \codett{Graph\_api} module.
 Given a term $t$, a class name $c$, and a path length $k$, it returns the names of the terms on a path no longer than length $k$ that belong to or subsumed by the class $c$.
 Then an analyst could use this function to define an analytic to find all people financially related to Sue Grafton who are associated with a person in a   criminal organization.
 Here is the OCaml code to create the analytic.
 \begin{verbatim}
 mk_analytic "find_criminal" "m_target" 
    "criminal_org" 
    (fun t -> nearest 3 t "criminal_org") 
 \end{verbatim}
 When Flutes DB runs analytics, it take all terms in the analytic's input class and passes it as an argument to the code representing the body of the analytic. If the code doesn't raise an exception during its execution, then FlutesDB checks if the output term can be subsumed by the output type.
If it can be, then its coerced term is returned as the result of the evaluation.

\section{Conclusion}

We demonstrated the benefits of using concepts from functional programming to create a language for extracting mission graphs from large graphs of linked objects. The proof-of-concept we presented in this paper, FlutesDB, can be used to create an end-user language for analysts because the user can focus on what needs to be solved and not on how to solve it. For example, the end-user does not have to provide the programming logic to perform graph transformations to extract mission graphs. The end-user only has to specify the characteristics of the mission graph. In cases where programming logic is needed, our proof-of-concept demonstrates programmers can easily provide analytics that analysts could incorporate into their specifications.

\subsection{Related Work}
Researchers have created approaches using SPARUL\ that can be used creating graph transformations\citeref{rak_making_2013}. However, these approaches have a fundamental problem they make them undesirable for intelligence analysis. The SPARUL and RDF type system is too weak for preventing careless mistakes that could be catastrophic for intelligence analysis. In particular, analyst can specify an object belongs to a type, but whose structure does not match the type. This weakness could lead to data integrity issues that could cause analysts to make incorrect predictions. 

Flutes is related to attributed type graphs with node type inheritance \citeref{lowe_algebraic_2015}. We know of no existing work on attributed type graphs with node type inheritance
that supports semantic relationships between attribute names. Attributed type graphs are defined as set-theoretic functions or using categories. Therefore, it is possible we can use attributed type graphs as a denotational model of Flutes.

Flutes is also related to declarative graph grammars\citeref{ebert_gretl:_2012,horn_graph_2015}. However we know of no  declarative graph grammars that support node type inheritance.

\section*{Acknowledgement}

The research presented in this paper was funded in part by ONR\ 
contract N00014-15-P-1172.

\bibliographystyle{plain}

%
\bibliography{../my-library}

\copyrightnotice

\end{document}